# WAKEFIELDS IN SUPERCONDUCTING RF CAVITIES AND THE IMPACT ON VUV FREE-ELECTRON LASER OSCILLATOR PERFORMANCE

Alex H. Lumpkin[1,#], Henry P. Freund[2,3], Matthias Reinsch[4,5], and Peter J.M. van der Slot[6]

[1]Fermi National Accelerator Laboratory, Batavia, IL 60510 USA

[2]Department of Electrical and Computer Engineering, University of New Mexico, Albuquerque, NM 87131 USA

[3]Institute for Research in Electronics & Applied Physics, University of Maryland, College Park, MD 20742 USA

[4]Lawrence Berkeley National Laboratory, Berkeley, CA 94720 USA

[5] now at the University of California (Berkeley), Berkeley, CA 94720 USA

[6]Faculty of Science and Technology, University of Twente, Enschede, the Netherlands

## ABSTRACT

The Fermilab Accelerator Science and Technology (FAST) facility is currently in operation with its linac based on TESLA-type superconducting rf cavities. Using a 3-MHz micropulse repetition rate with  a long macropulse composed of up to 3000 micropulses, and with beam energies demonstrated at 300 MeV and projected to reach 800 MeV with two additional cryomodules, the feasibilies for a vacuum ultraviolet (VUV) and an extreme ultraviolet (EUV) free-electron laser oscillator (FELO) with the two energies are evaluated. We have used both the GINGER code with an oscillator module and the MINERVA/OPC code to assess FELO saturation prospects at 120 nm with a 5.0-cm-period undulator of 4.5-m length and the MINERVA/OPC code to assess the FELO at 13.4 nm with adjusted parameters. The simulation results support saturation at both of these wavelengths which are much shorter than the demonstrated shortest wavelength record of 168.6 nm from a storage-ring-based FELO. This indicates superconducting rf linac-driven FELOs can be extended into this VUV-EUV wavelength regime previously only reached with single-pass FEL configurations. In addition, emittance-dilution effects due to wakefields in the cavities and the resulting sub-macropulse centroid slew effects on FELO performance are addressed using MINERVA/OPC simulations for the first time.

Index: 41.60

[#]lumpkin@fnal.gov

## I. INTRODUCTION

Over the last few decades since the initial description of the free-electron laser (FEL) concept [1] and the subsequent demonstrations first as amplifiers [2, 3] and then as oscillators at infrared wavelengths [4-8], the challenges of generating shorter wavelengths have been related to electron beam quality, beam energy, available undulators, and available optical cavity mirrors of high reflectivity. Over a 15-year period starting in the late 1980s, storage-ring-based FEL oscillators (FELOs) [9-11] used the improvement of mirror reflectivities at VUV wavelengths to push the shortest wavelength record below 200 nm to 176 nm [12] and recently to 168.6 nm [13]. Roughly in parallel, the issue of low mirror reflectivities was avoided by moving to single-pass FELs based on self-amplified spontaneous emission (SASE). These exploited the much higher beam brightness offered by photo-injected normal conducting linacs [14,15] and superconducting linacs [16] as shown in the last decade in the visible, EUV, and soft x-ray regimes. Now, SASE FEL configurations have been successfully extended to the hard x-ray regime with normal conducting radio frequency linacs [16-20] and with superconducting radio frequency linacs [21,22].

Free-electron laser oscillators typically rely on low-gain undulators in combination with high-$Q$ resonators such as, for example the high average power infrared FELO experiment [23] at the Thomas Jefferson National Accelerator Facility. In contrast, concepts have also been discussed for using a high-gain undulator within a low-$Q$ resonator [24], and this configuration has been termed a regenerative amplifier (RAFEL). Concepts for hard x-ray FELOs [25,26] and RAFELs [27,28] have been discussed. Proposals for attaining transform-limited EUV light in an oscillator configuration based on a ring resonator using multi-faceted mirrors with a normal conducting accelerator by the Los Alamos National Laboratory group [29] were developed, and the oscillator configurations were subsequently overshadowed by the SASE FEL successes in the EUV.

In this context, a significant opportunity exists to enable the first vacuum ultraviolet (VUV) and extreme ultraviolet (EUV) FELO experiments at the Fermilab Accelerator Science and Technology (FAST) facility [30,31]. The bright beam from the L-band photo-injector would provide sufficient gain/pass to compensate for reduced mirror reflectances in the VUV-EUV regimes, the 3-MHz micropulse repetition rate for up to 1 ms will support an oscillator configuration, the superconducting RF linac (SCRF) will provide stable energy, and the possible GeV-scale energy with three TESLA-type cryomodules will satisfy the FEL resonance condition in the EUV regime. Concepts based on combining such beams with a 5-cm-period undulator and with an optical resonator cavity in an FEL oscillator configuration are described. We used the 80% reflectances at 120 nm [32] and 68% reflectances at 13.4 nm for normal incidence on multilayer metal mirrors developed at Lawrence Berkeley National Laboratory (LBNL) [33]. Simulations using GINGER [34] with an oscillator module and MINERVA/OPC [35 – 39] show saturation for a 120-nm case after 80-140 passes, and for a 13.4-nm case MINERVA/OPC predicts saturation after 250 passes. Initially, VUV experiments would begin in the 160 – 120 nm regime with beam energies of 260 – 300 MeV. These latter electron beam energies have already been demonstrated with the injector and first cryomodule [40]. In addition, we describe fundamental results on emittance dilution and submacropulse beam centroid slewing and oscillations due to wakefields in TESLA-type cavities [41-43] that must be controlled for FELO optimization as guided by MINERVA/OPC simulations at 120 nm and 13.4 nm. The high gain per pass enables the FELO operation with hole outcoupling while avoiding the stringent mirror-reflectance requirements of the storage-ring-based FELOs.

## II. THE FAST FACILITY

The FAST linac includes an L-band photocathode (PC) rf gun, two booster L-band SCRF accelerators (denoted CC1 and CC2), the chicane bunch compressor, and the first cryomodule as schematically shown in Fig. 1. The PC gun uses a $Cs_2Te$ photocathode which is irradiated by the ultraviolet component of the drive laser with a 3- or 9-MHz micropulses-repetition rate as described elsewhere [44]. The L-band accelerating sections provide 40- to 50-MeV beams before the chicane, and an additional acceleration capability up to a total of 300 MeV is installed and commissioned [40]. Additional acceleration capability could potentially be installed in the form of three cryomodules total with eight 9-cell cavities (an International Linear Collider (ILC) rf unit). The higher-order mode (HOM) coupler locations are indicated in the cryomodule and the HOM detectors for the first two dipolar mode passbands from 1.6 to 1.9 GHz were described previously [41]. The first cryomodule is presently installed and those single cavities have been conditioned to an average of 31.5 MV/m [40] and thereby met the ILC gradient target. Additionally, the proposed FELO cavity resonator area in the high-energy end is indicated in Fig.1 and shown in Fig.2.

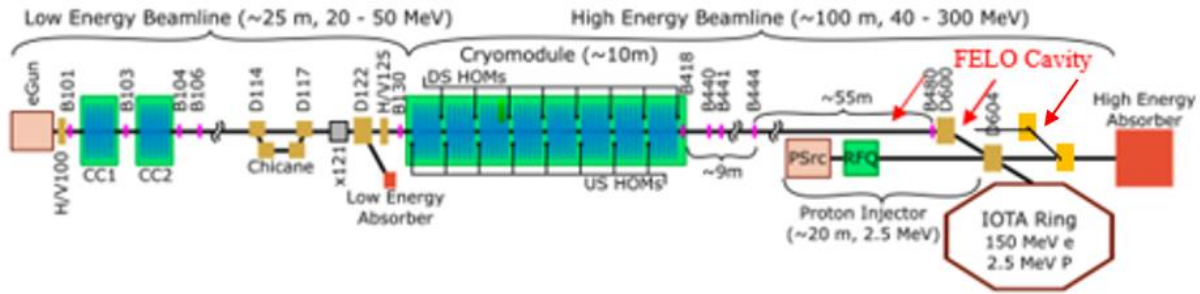

Fig. 1: Schematic of the FAST facility showing the PC rf gun (eGun), booster accelerators CC1 and CC2, chicane, the rf BPMs denoted as Bnnn, the X121 imaging station, the dipoles denoted as Dnnn, first cryomodule, and the proposed location of the FELO resonator (see Fig. 2).

At the end of a 1-m drift located 10 m downstream of CC2, the X121 imaging station provides transverse emittance data based on a quadrupole field scan, the YAG:Ce screen, and the Prosilica digital CCD camera images. Additionally, the phase of the CC2 section can be adjusted to energy chirp the beam entering the first chicane to vary bunch-length compression. The longitudinal distributions are evaluated with an optical transition radiation (OTR) converter screen of aluminized Si selectable by a stepper actuator at the same X121 imaging station. We use an all-mirror optical transport to a Hamamatsu C5680 streak camera with synchroscan vertical unit phase locked to 81.25 MHz [42]. Streak camera measurements have been done for various micropulse charges and with various compressions up to 5 in the first chicane with 40 MeV achieved [45]. Space-charge forces in the gun result in longer bunches at higher charges.

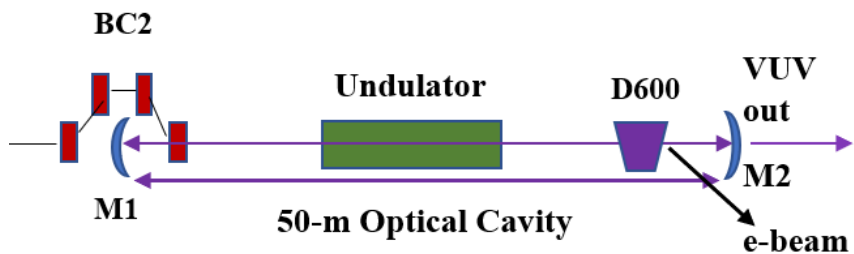

Fig. 2: Schematic of a potential configuration for the FELO in the high energy transport area. A chicane would be added to provide access for the upstream mirror of the resonator. The e-beam enters from the left and is directed off axis before the M2 mirror by the dipole D600.

Table 1: Summary of demonstrated electron beam properties at FAST.

| Parameter | Units | Value |
|---|---|---|
| Micropulse Charge | pC | 500 |
| Emittance, norm. | mm-mrad | 2 – 3 |
| Bunch Length (σ) | ps | 2 |
| Micropulse Number | | 1 – 3000 |

A micropulse charge of 500 pC will be used typically as indicated in Table 1. The nominal micropulse format is 3 MHz for up to 1 ms which is unique for test facilities in the USA and highly relevant to the next generation of FELs. The macropulse repetition rate will be 5 Hz. Simulations of the photoinjector with ASTRA [46] and the transport dynamics through the cryomodule and to a second chicane have been previously reported. With the bunch compression spread over two stages the emittance growth is controlled and at 250 pC one still obtains the final emittances less than 2 mm mrad for a 1-ps pulse length [47]. These start-to-end simulations were the basis of the electron beam properties invoked for the FELO simulations reported in a later section. The third harmonic linearizer and the second chicane are proposed to support much higher peak currents of 1100 A for the 13.4-nm case.

## III. THE U5.0 UNDULATOR

The propagation of the electron beam through an undulator results in the generation of photons. This is initiated through the spontaneous emission radiation (SER) process, but under resonance conditions a favorable instability evolves as the electron beam co-propagates with the optical field, and the electron beam is microbunched at the resonant wavelength. In the oscillator configuration, the subsequent passes of the next e-beam micropulses with the photon beam in the resonant cavity can lead to saturation. For a planar undulator, the resonant wavelength is given by $\lambda = \lambda_u (1 + K^2/2)/2n\gamma^2$, where $\lambda$ is the output wavelength, $\lambda_u$ is the undulator period, $K$ is the undulator field strength parameter, $n$ is the harmonic number, and $\gamma$ is the relativistic Lorentz factor. The U5.0 undulator proposed for the studies has been transferred from Lawrence Berkeley National Laboratory (LBNL) to Fermi National Accelerator Laboratory (FNAL) after retirement from the Advanced Light Source (ALS) storage ring. It is 4.5-m long with an undulator period of 5.0 cm. [48]. It has stepper motor control of the magnetic gap to provide a tunable $K$ value. This feature is a strong advantage for our applications at FAST. A summary of the key parameters is provided in Table 2.

Table 2: Summary of the U5.0 undulator parameters [46].

| Parameter | Value | Units |
|---|---|---|
| Period, $\lambda_u$ | 5.0 | cm |
| Length in Periods, $N$ | 89 | |
| Length, $L$ | 4.55 | m |
| Maximum Field at 1.4 cm | 0.89 | T |
| Magnetic Gap Range | $1.4 - 2.17, 4.7$ | cm |
| Harmonics | 3, 5 | |
| Range of $K$ | $0.45 - 3.9$ | |

## IV. SIMULATION CODES

Two simulation codes were used to model various FEL options at different wavelengths and beam energies: GINGER and MINERVA/OPC. GINGER [34] was originally developed to simulate FEL amplifiers, but it has been extended by inclusion of an oscillator module. MINERVA, a successor to MEDUSA [35], was also originally developed to simulate amplifier and (SASE) FELs [35,36], and was subsequently linked with the Optical Propagation Code (OPC) [38] to model FEL oscillators [37]. Both GINGER and MINERVA employ the slowly-varying envelope approximation, but the similarity ends there.

GINGER is a multidimensional ($r$-$z$-$t$ fields, $x$-$y$-$z$-$t$ macroparticles), polychromatic FEL simulation code developed over more than 25 years. GINGER also supports monochromatic simulations, meaning that all field quantities (and many particle quantities such as the particle bunching) vary exactly as exp ($-i\omega t$). Other quantities such as beam current and energy are presumed to be approximately time-invariant over "slow" timescales (i.e., when averaged over ~ dozens of wave periods). The code models shot noise, slippage, current and energy variations. The code has a polychromatic oscillator mode.

MINERVA is a fully three-dimensional, time-dependent simulation code that includes harmonics and start-up from noise. It models helical, planar, and elliptical wigglers and the optical field is represented as a superposition of Gaussian modes. Electron trajectories are integrated using the three-dimensional Lorentz force equations in the combined magnetostatic and optical fields. No wiggler-averaged orbit analysis is used so that all harmonic contributions to the electron orbits are self-consistently included. Models for quadrupoles and dipoles are included. In time-dependent mode, the electron bunch and the optical pulse are described by an ensemble of temporal slices where each slice is advanced from $z \rightarrow z + \Delta z$ as in steady-state simulations, after which the field is allowed to slip relative to the electrons. Since MINERVA must resolve the wiggler-motion in the undulator(s), typically $\Delta z << \lambda_u$, and slippage is interpolated over this distance at the overall rate of one wavelength per wiggler period. In simulating an FEL oscillator, MINERVA treats the propagation of the optical pulse through the undulator and then maps the Gaussian modes onto the grid used by OPC which is then passed to OPC. OPC then propagates the optical field using either the Fresnel diffraction integral or the spectral method in the paraxial approximation using fast discrete Fourier transforms (FFT). A modified Fresnel diffraction integral is also available and allows the use of FFTs in combination with an expanding grid on which the optical field is defined. Currently, OPC includes mirrors, lenses, phase masks, and round and rectangular diaphragms. Several optical elements can be combined to form more complex optical components, e.g., by combining a mirror with a hole element, extraction of radiation from a resonator through a hole in one of the mirrors can be modeled. Phase masks can be used to model mirror distortions or to create non-standard optical components like a cylindrical lens. After propagating the optical field through the resonator and back through the undulator, OPC then passes the optical field to a utility program (MERCURY) which decomposes the field on the grid into Gaussian optical modes which is then used to restart MINERVA for another pass through the undulator. This procedure is repeated for any desired number of passes. The MINERVA/OPC procedure has been validated by comparison with the 10-kW Upgrade Experiment at Jefferson Laboratory [37]. We note that previous considerations of a hole-outcoupled FELO by Prazeres et al. [49] obtained an off-axis laser alignment within the optical cavity due to the mode's avoiding the extraction hole, but we use axially symmetric electromagnetic fields in our MINERVA simulations.

## V. SIMULATION RESULTS

Following single-pass experiments that will be used for non-intercepting electron beam diagnostics [50], an intriguing application is the investigation of VUV-EUV FELO configurations. The FAST pulse train at 3 MHz (or 9 MHz) provides the enabling technology, and it is characterized by a bright electron beam with a nominal transverse emittance of 2 mm-mrad, peak currents of 100 A, potential GeV-scale energies, and energy spreads of $5 \times 10^{-4}$. Numerical studies using a ring resonator optical configuration were executed at Los Alamos National Laboratory with the 3D FEL simulation code, FELEX, in the VUV and XUV regimes [51]. At this time at FAST, one can expect to surpass their emittance speculations, albeit with lower charge per micropulse. Although they considered a 1.6-cm-period undulator with a low $K$ value, we can use higher energies than their 261 MeV to reach the resonance conditions in the $40 - 50$ nm regime with only two cryomodules installed. Initial gain length evaluations could be done empirically in the single-pass mode before the final design, and a test of the resonator optical path tuning could be initiated with UV-Visible light with only the present cryomodule operating.

Base FELO options are summarized in Table 3. Phase numbers 1-3 are basically related to the number of installed cryomodules with up to 250 MeV beam acceleration per cryomodule. Our nearest term cases are the phase 1 entries with energies from 125 to 300 MeV. They were evaluated with the U5.0's 5-cm period in the table, but we note that much shorter wavelengths might be generated, in principle, with 300 MeV with for example the 1.8-cm period, 3.56-m-long device at Argonne National Laboratory [52].

Table 3: Summary of possible FELO wavelengths generated with a 5.0-cm period undulator at FAST. The Phase numbers represent the number of installed cryomodules with standard gradient. Phase 1 is installed.

| Phase # | Beam Energy (MeV) | FEL Fund. (nm) | Period (cm), $K$ | FEL Harmonics (nm) 3, 5 |
|---|---|---|---|---|
| 1 | 125 | 680 | 5.0, 1.2 | 226 |
| 1 | 150 | 472 | 5.0, 1.2 | 157 |
|   | 200 | 265 | 5.0, 1.2 | 88 |
| 1 | 250 | 170 | 5.0, 1.2 | 57 |
| 1 | 250 | 262 | 5.0, 1.8 | 87 |
| 1 | 300 | 120 | 5.0, 0.8 | 40 |
| 2 | 500 | 42 | 5.0, 1.2 | 14, 8.3 |
| 3 | 800 | 16 | 5.0, 1.2 | 5.3 |
| 3 | 800 | 13.4 | 5.0, 0.9 | 4.4 |

Our interest in the use of the U5.0 device was solidified using simulations of a FELO performance at various wavelengths. Initial simulations were performed at LBNL using the GINGER code [34] with an oscillator module and assessed the initial target of 120 nm, using nominal expected electron beam parameters. Subsequently, the MINERVA/OPC [35-39] simulations were performed to extend our studies and explore cavity and hole out-coupling. A comparison of the code results at 120 nm is presented below. With the lower reflectance of the VUV-EUV mirrors, the gain per pass and losses are critical considerations. The high brightness beam produced by the photoinjector, and the long pulse train, are key features in the build up to saturation. The nominal cavity length of 50.0 m is invoked to match the 3-MHz micropulse repetition rate to the optical cavity roundtrip time. The experimental program should provide an ideal benchmarking of the simulations. The electron beam parameters and resonator aspects used are given in Table 4.

Table 4: Summary of nominal electron beam parameters and resonators used in the simulations.

| Parameter | 120-nm Case | 13.4-nm Case |
|---|---|---|
| Energy | 300 MeV | 800 MeV |
| Normalized Emittance | 2.0 mm-mrad | 2.0 mm-mrad |
| Peak Current | 100 A | 1100 A |
| Resonator optics | Concentric,1-mm hole | Concentric, 0.3-mm hole |
| Mirrors | Amorphous C on Si | Mo/Si multilayer |

**A.  Simulated Operation at 120 nm**

In this section, we discuss simulation of 120-nm FELOs using both GINGER and MINERVA/OPC. The microbunching and power evolution of a FELO operating at 120 nm driven by a 300-MeV electron beam with $I_{peak}$ of 100 A are shown in Fig. 3 as simulated with GINGER using an out-coupling hole of 1-mm radius. GINGER indicates a strong microbunching fraction in the fundamental and third harmonic as shown at the left. Power saturation is reached after only 120 passes at 100 MW at the end of the undulator, as shown on the right. The fundamental bunching fraction, which is the fraction of the beam modulated at the resonant wavelength, is calculated to be ~ 0.35 (with the third harmonic bunching fraction calculated at 0.15). With a mode size of about 4 mm we would outcouple about 5 MW per pass in the fundamental at saturation. It is important to observe that the GINGER results indicate power at the undulator exit, while MINERVA/OPC describes the power at both the undulator exit and the resonator exit.

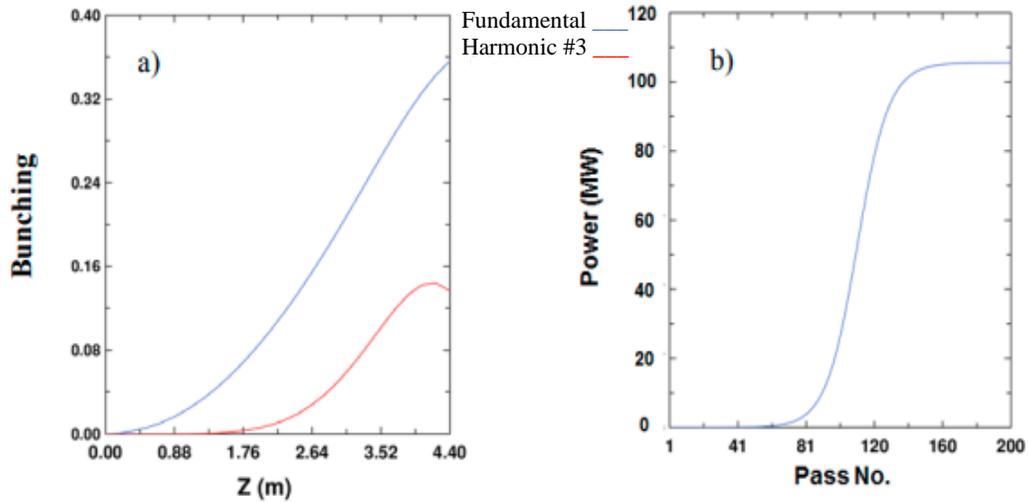

Fig. 3: Simulations of the 120-nm FELO a) bunching fraction for the fundamental (blue) and 3rd harmonic (red) and b) power saturation using GINGER with 1-mm radius hole outcoupling.

Table 5: Parameters used in the MINERVA simulations.

|  | 120-nm Case | 13.4-nm Case |
|---|---|---|
| **Electron Beam** | | |
| Kinetic Energy | 301.25 MeV | 804.0 MeV |
| Peak Current | 100 A | 1100 A |
| rms Energy Spread | 0. 05% | 0.15% |
| Normalized Emittance | 2.0 mm-mrad | |
| Beam Size $x$ (rms) | 122 μm | 99 μm |
| Beam Size $y$ (rms) | 120 μm | 89 μm |
| Twiss $\alpha_x$ | 1.0 | 2.0 |
| Twiss $\alpha_y$ | 1.0 | 1.5 |
| Repetition Rate | 3.0 MHz | |
| **Undulator** (flat pole face) | wiggle plane in $x$ | |
| Period | 5.0 cm | |
| Magnitude | 2.4614 kG | 1.701 kG |
| Length (1 period entry/exit taper) | $90\lambda_w$ | $130\ \lambda_w$ |
| **Resonator & Optics** | concentric, hole out-coupling | |
| Cavity Length | 50 m | |
| Mirror Curvature | 25.32 m | 25.09 m |
| Rayleigh Range | 2.35 m | 0.75 m |
| Hole Radius (downstream mirror) | 1.0 mm | 0.3 mm |

The parameters for the MINERVA/OPC simulations of both the 120 nm and 13.4 nm FELOs are shown in Table 5. Simulations with MINERVA/OPC yielded a saturated peak power at the undulator exit of 152 MW and output power levels of 7.8 MW after 60 – 70 passes as shown in Fig. 4 using a nominal 2-mm-mrad beam emittance, a 0.05% rms beam energy spread with a 301 MeV/100 A beam, 80% reflective mirrors, and a 1-mm radius output coupling hole. The predictions from GINGER and MINERVA/OPC substantially agree. The two codes agree for the power of ~100 MW at the undulator exit, although MINERVA/OPC calculates fewer passes to saturation at ~80. Such a demonstration would shatter the existing FELO shortest wavelength record at 168 nm on the fundamental done in a storage ring [13]. At ~300 MeV this could be done with the injector plus the one cryomodule that is presently installed. The eight cavities have achieved the average of 31.5 MV/m in a full beam test [31,40].

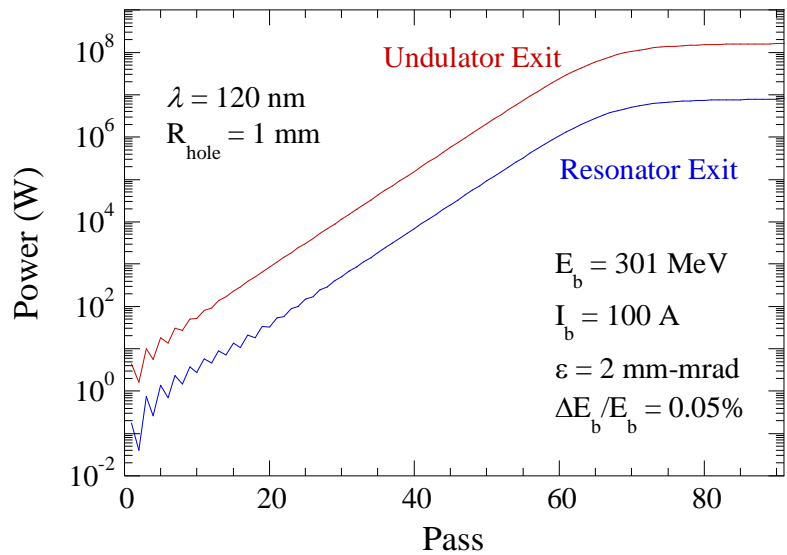

Fig. 4: Simulations of the 120-nm FELO power saturation using MINERVA/OPC with 1-mm radius hole out-coupling in the downstream mirror with nominal FAST electron beam parameters.

We also considered the critical aspect of the power and energy densities on the resonator mirrors. The results of MINERVA/OPC simulations for the downstream mirror are shown in Fig. 5. For the 120-nm case, we calculated 500 MW/cm$^2$ per pulse maximum on the downstream mirror where hole outcoupling with a 1-mm radius proved advantageous by avoiding higher densities on axis (see Fig. 5). Using a bunch length of 2 ps (rms), we estimate the maximum energy density at 1.0 mJ/cm$^2$ on the downstream mirror and ~1.4 mJ/cm$^2$ on the upstream mirror on axis, far below the damage threshold value of 70 mJ/cm$^2$ reported at 98 nm and normal incidence for amorphous carbon on Si substrates [53], using an FEL in SASE mode operated with up to 100s of micropulses.

This mirror type thus would be a credible candidate for us since we would expect a similar performance at 120 nm.

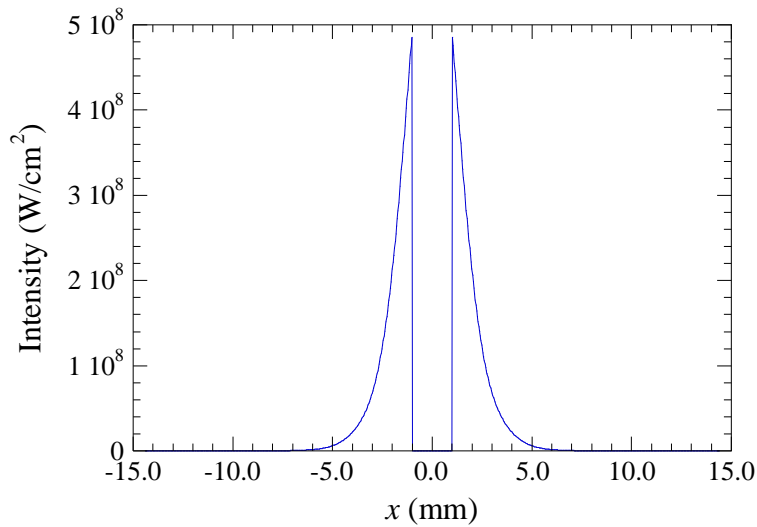

Fig. 5: MINERVA/OPC simulation of the power density on the downstream mirror for the 120-nm case with mirror reflectivities of 80% and a 1-mm hole radius.

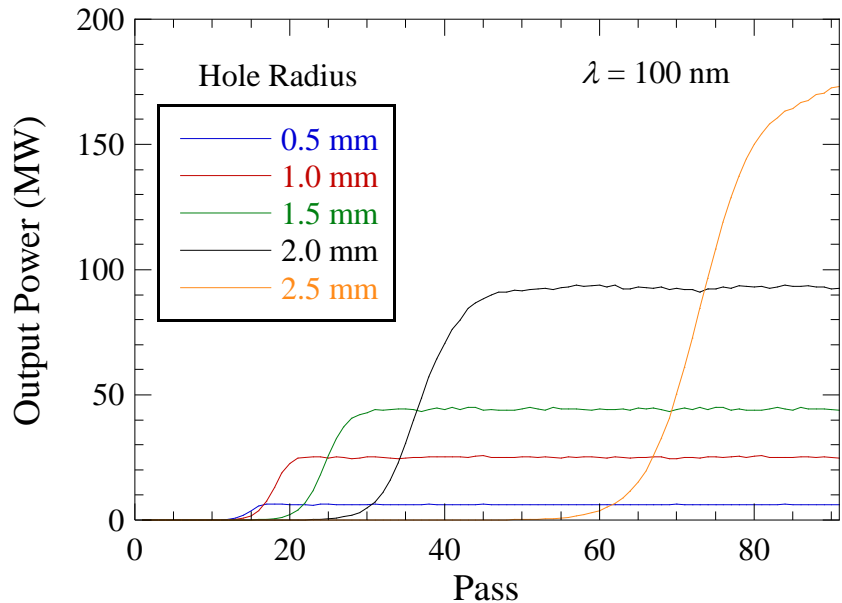

Fig. 6: MINERVA/OPC simulations at 100 nm showing the out-coupled power vs pass for a variety of hole radii.

## B. Simulated Operation at 100 nm

A separate assessment of the hole out-coupling was performed with MINERVA/OPC at 100 nm with a 501-MeV electron beam as shown in Fig. 6 for a variety of hole radii. The full dynamic treatment of the cavity modes in the oscillator requires a full time-dependent simulation which is beyond the scope of the present work. Such simulations require the inclusion of multiple temporal slices of the electron bunch and optical field which markedly increases the run times.

The MINERVA/OPC formulation in the steady-state regime implicitly assumes the ideal cavity synchronism; however, in time-dependent simulations the cavity length must be chosen to correctly model the cavity detuning which describes the overlap between the returning optical pulses at the undulator entrance with the incoming electron bunches. This necessitates many separate multi-pass simulations which is a very time-consuming aspect. However, the MINERVA/OPC formulation is capable of modeling this and implicitly includes a treatment of the limit cycle oscillations which is dependent upon the cavity detuning. Figure 6 shows the power out-coupled from the resonator versus pass number for five different hole radii: 0.5 mm, 1.0 mm. 1.5 mm, 2.0 mm, and 2.5 mm. The hole size affects the out-coupled power in oscillators which is a balance between increasing the fraction of the out-coupled power which must be balanced by the gain in the undulator. Typically, oscillators saturate when the gain in the nonlinear regime drops to the level of the loss. As the loss increases with increasing hole size, it reaches a point where the small-signal gain cannot overcome the losses and the oscillator cannot lase. In the current example, this is reached when the hole radius exceeds 2.5 mm. Because the net gain decreases as the out-coupled power increases, the number of passes required to reach steady state increases with the hole size as well. This is evident from the figure.

## C. Simulated Operation at 13.4 nm

The bright FAST beam allows running with lower reflectivity mirrors. Absorption of power is an issue that can be addressed by mirror cooling techniques and hole outcoupling. The schematic of the oscillator to be located in the high-energy beamline of the FAST facility was shown in Fig. 2. The upstream and downstream mirrors will be 50.0 m apart to provide a roundtrip optical path of 100 m that matches the electron beam micropulse 333-ns spacing for a 3-MHz rate. A chicane will allow the positioning of the upstream mirror on axis. Transport of the VUV radiation will initially be done to local diagnostics, but eventually it could be transported to the upstairs optics lab or a downstream lab. A more challenging short-wavelength regime at 13.4 nm was also evaluated. In these cases, a longitudinal phase space linearizer based on a third harmonic rf cavity (CAV39), a second compression (BC2), and two additional cryomodules would be needed as shown in Fig. 7 [47].

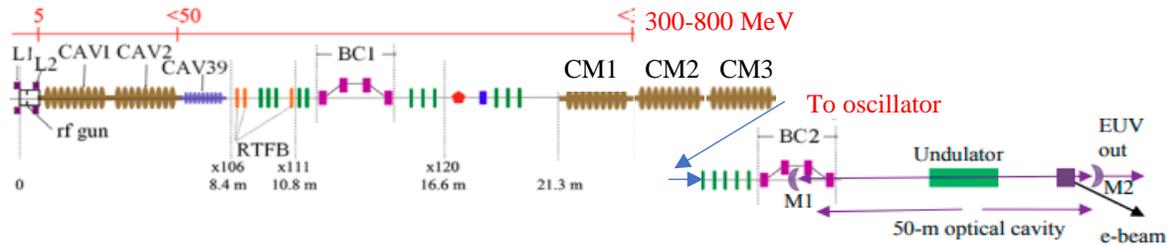

Fig. 7: Schematic of extended capability facility for short wavelength experiments at 13.4 nm with the rf gun, capture cavities 1 and 2, the third harmonic cavity, BC1, three cryomodules (CM1-3), matching quadrupole elements, BC2, the undulator before dipole D600, and the optical resonator cavity mirrors, M1 and M2.

We note that recent investigations with modified low-temperature bake protocols demonstrated gradients of 50 MV/m in a TESLA shaped cavity [54] so one could envision only needing two cryomodules with eight such cavities each to reach the target 800 MeV. Operating at shorter wavelengths involves the use of lower reflectivity mirrors, even when using hole out-coupling.

We assume a mirror reflectivity (R) of 68% in line with proven mirror technology [33,55]. This means that the total mirror loss $L = (1 - R^2)$ is about 54% not counting the loss through the hole which increases the total loss. Since the oscillator in the steady state will require a gain $G = L/(1 - L)$, this means that the saturated gain must be in excess of about 130%. However, this is approximately the small-signal gain of this system using the 4.5 m undulator with a peak current of 900 A. Therefore, the MINERVA/OPC simulations use a higher peak current of 1100 A and an undulator with a length of 6.5 m at a beam energy of 800 MeV to increase the small-signal and unsaturated gains. We remark here that the losses from the hole can have a marked effect on the lasing, and we find that the largest hole radius consistent with lasing with these beam and undulator parameters is 300 μm.

The basic parameters of the MINERVA/OPC simulations are listed in Table 5 for the FELO at 13.4 nm using the 5.0-cm period undulator of Table 2 but now lengthened to 6.5 m. Note that the beam has been matched into the undulator so that the beam waist is near the center of the undulator, as shown in Fig. 8. It should be remarked here that Fig. 8 exhibits the evolution of the beam envelope relative to the centroids ($x_c$ and $y_c$) in the $x$- and $y$-directions which is primarily governed by the choice of the initial Twiss parameters which determine the focusing of the beam.

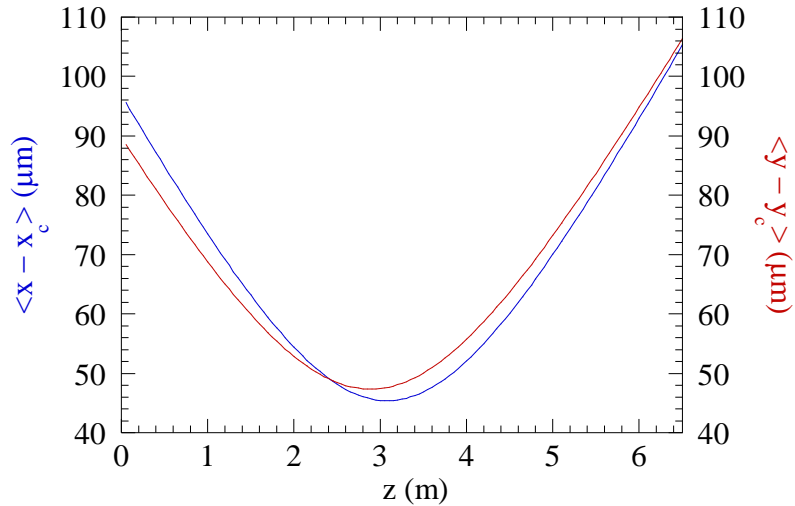

Fig. 8: Plot of the $x$ match (blue) and $y$ match (red) of the beam into the undulator for the 13.4- nm case. This denotes the beam envelope about the centroid.

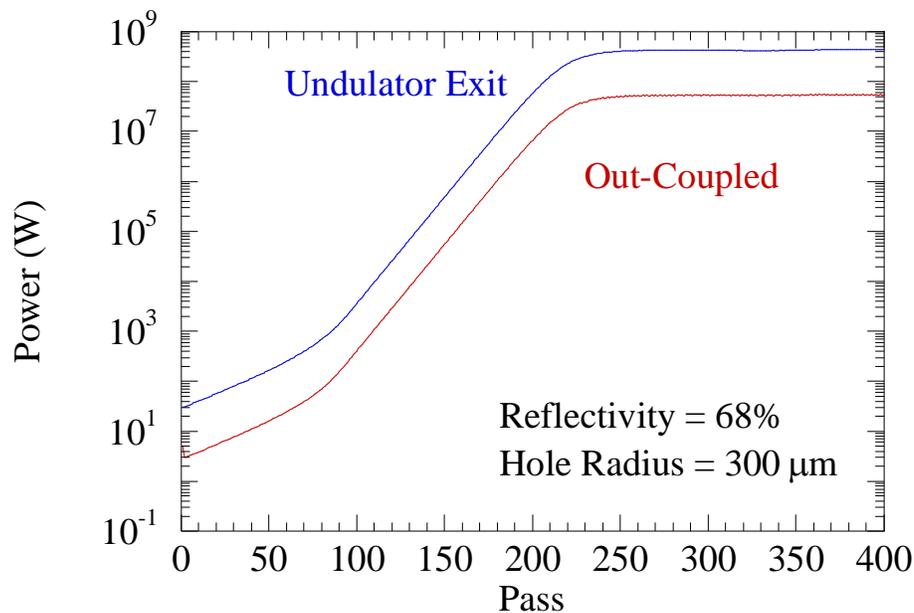

Fig. 9: MINERVA/OPC simulation results at 13.4 nm with mirror reflectivities of 68% and a 300-μm hole radius.

The MINERVA/OPC simulations made use of a 300-μm hole radius with a mirror reflectivity of 68% as shown in Fig. 9. It is evident that the steady state is achieved after about 250 passes and that the power at the undulator exit is 332 MW and the out-coupled power is 52.9 MW. These are very encouraging results invoking the reflectivities of the present generation multilayer metal mirrors optimized at this wavelength and showing saturation is possible. Other optical resonator configurations and mirror cooling may be needed to address the heat loading and consequent thermal distortions of the mirrors [25]. We also did not handle the details of optical phase shifts in narrow band multilayer mirrors such as reported in proposed hard x-ray FELOs [36] with the steady-state codes.

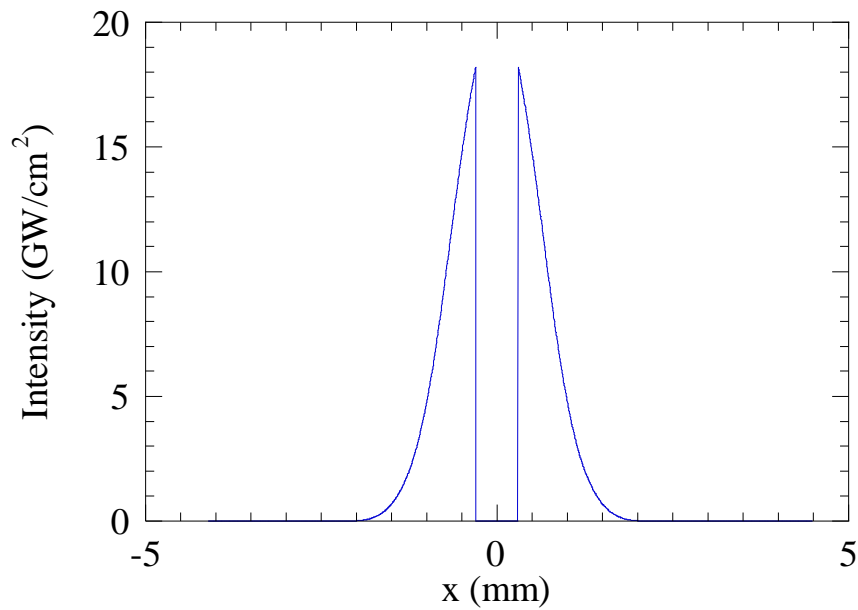

Fig. 10: MINERVA/OPC simulations of the power density on the downstream cavity mirror for the 13.4 nm case with a 300-μm hole radius.

We again considered the critical point of energy density on the mirrors as calculated in simulations for the downstream mirror shown in Fig. 10. The multiplication of the peak power density of 18 GW/cm$^2$ in the downstream mirror by an rms bunch length of 0.2 ps gives 3.6 mJ/cm$^2$, well below the damage threshold reported [55] of 83 mJ/cm$^2$ at 13.5 nm for a Mo/Si multilayer mirror. This FLASH test used 400 SASE FEL micropulses at a 1-MHz repetition rate with a 10-Hz macropulse rate. The peak intensity on axis on the upstream mirror is higher at about 250 GW/cm$^2$ corresponding to an energy density of 50 mJ/cm$^2$, which is still below the reported damage threshold.

## VI. WAKEFIELD EFFECTS ON BEAM AND FELO DYNAMICS

Since our initial considerations of the FELO [56], we have identified and mitigated potential long-range wakefields (LRWs) [41,43] and short-range wakefields (SRWs) [42] generated by off-axis beam transiting the TESLA-type Superconducting rf cavities in FAST. These beam-dynamics effects that include submacropulse electron beam centroid slewing and oscillations and submicropulse head-tail kicks, respectively, would directly affect the FELO performance that has been simulated. We briefly describe the empirical evidence [41-43] for wakefield effects and the evaluations of the simulated effects of such dynamics on the FELO.

### A. Empirical Results

The initial investigations on the LRWs became focused on the higher-order modes (HOMs) generated in TESLA-type cavities as a source of submacropulse effects. In particular we studied the dipolar HOMs that might couple to the beam most strongly in the first two passbands from 1.6 to 1.9 GHz. Our investigations started on the two single cavities CC1 and CC2 (see Fig. 1). We showed that an observed vertical beam centroid oscillation with a 100-kHz difference frequency in the downstream rf BPMs (with bunch-by-bunch capability) could be explained by a near-resonant effect of the vertical polarization component of dipolar mode 14 in CC2 and a beam harmonic [41]. This oscillation amplitude grew to ~500 µm in a 10-m drift after this cavity. An example of the beam vertical oscillation and slew over 500 bunches is shown in Fig. 11a and mitigation by steering with corrector V101 in Fig. 11b for 50 bunches. We saw no corresponding effect in the horizontal plane in CC2 due to the fact that the actual frequency of the horizontal polarization component was not near enough to a beam harmonic. We identified horizontal centroid slew and oscillation from the CC1 ascribed to its dipolar modes 7 and 30 induced by horizontally steering the beam before the cavity.

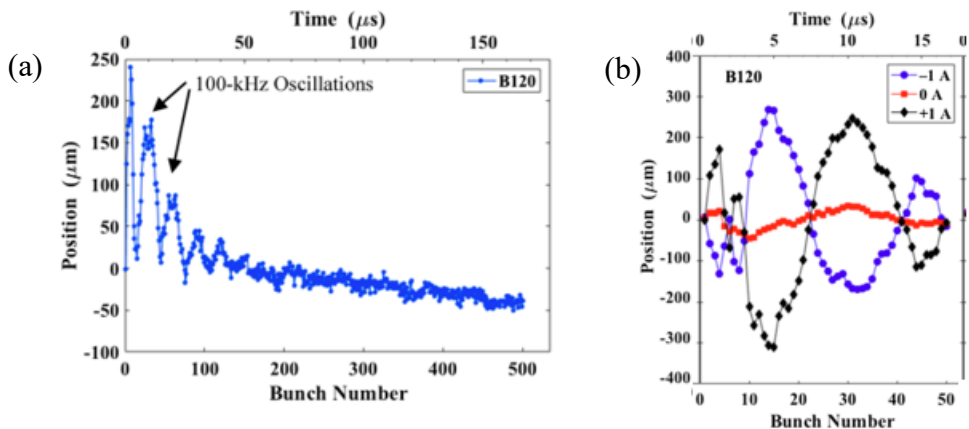

Fig. 11. (a) Vertical centroid oscillations shown at rf BPM location B120 after CAV2 for 500 b, 500 pC/b, and V101= +1 A. The 100-kHz oscillation decays noticeably in the first 200 b, and a centroid slew continues to the end of the macropulse. (b) Submacropulse effects in the first 50 bunches with corrector optimization at a reference current of 0 A (red curve) [41].

We next investigated SRWs in the single cavities. We were able to show that the magnitude of the head-tail kick of about 100 µm that would occur in each micropulse for a 1-mm beam offset from the cavity center was larger than the HOM effects seen in the centroid slew [42]. We were also able to detect projected vertical profile changes of 50% as shown in Fig. 12a (-0.5-A vs +2.0 A curves) and to generate head-tail kick magnitudes of ~200 µm correlated with the steering (and HOM signals) as shown in Fig. 12b [57]. The charge dependence is also seen as the kick magnitude increases with higher charges for both off-axis steering directions in Fig. 12b. Of course, both the HOM kicks and SRWs depend inversely on beam energy so on-axis steering is more critical in the early cavities. Cavity misalignments and the near-resonance effects with beam harmonics are a major part of the two stories.

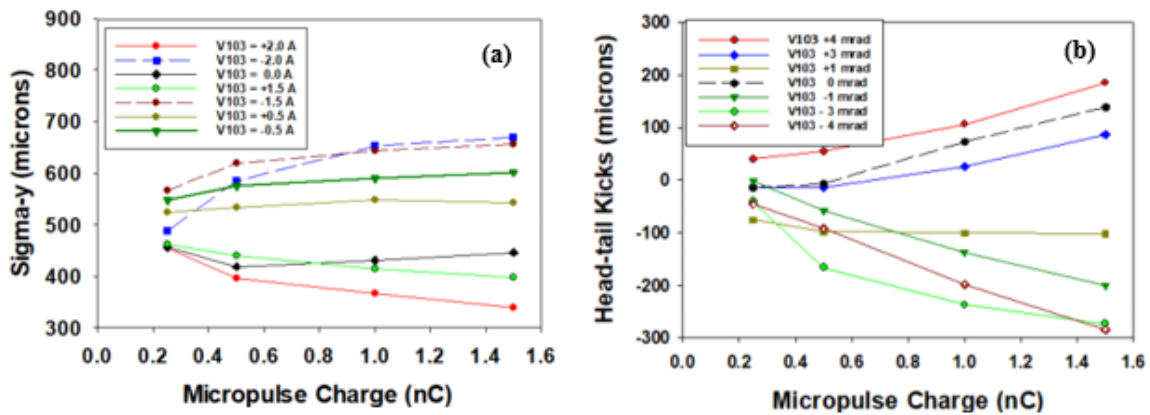

Fig. 12: (a) Projected vertical beam profiles for X121 streak camera images versus charge at various V103 corrector values. (b) Observed head-tail centroid kicks for the same images as a function of charge and V103 corrector values. A 1-A corrector current change corresponds to a 2-mrad angular steering change into CC2 [57].

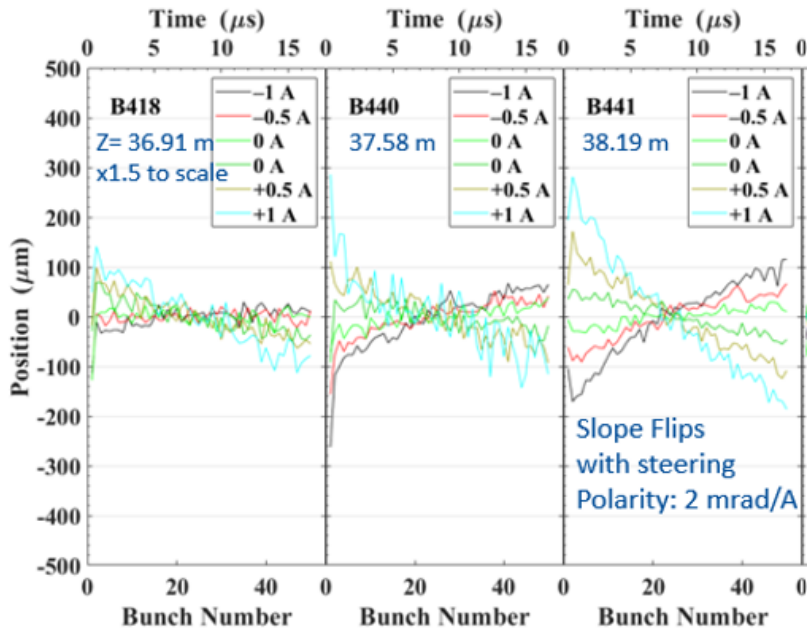

Fig. 13: Observation of the V125 corrector-correlated, vertical beam-centroid oscillation and residual slewing at B418, B440, and B441 with z positions noted. The corrector magnet steering effect corresponds to ~ 2 mrad/A [43].

More recently, we have applied our techniques on correlating HOM signal strength and beam dynamics effects in a full cryomodule at FAST [43]. We observed significant submacropulse centroid slewing in the first 50 bunches at the three 3 rf BPMs located after the cavity C8, particularly 1.2 m after B441 in Fig. 13, by steering the beam with corrector magnets located 4 m before the entrance of the cryomodule. In addition, we observed a submacropulse beam centroid slew of ~300 µm and a 240-kHz oscillation 68 m downstream for 2-mrad input steering. These tests were done with a 25-MeV input energy and 100-MeV output energy and 125 pC per bunch. We note that the wakefield effects increase with charge linearly so at higher charges of 500-1000 pC this aspect will compensate for their inverse dependence on the higher exit beam energy of 300 MeV.

## B. Simulations of Beam dynamics from Wakefields on the FELO Performance

The MINERVA/OPC simulations were done in the steady-state regime with only a single temporal slice so there is no issue with the cavity detuning. The MINERVA simulations start with shot noise so that no initial seed is provided [36, 58-60]. The basic parameters of the simulations are listed in Table 5 for the FELO at 120 nm using the same 5.0-cm period undulator of Table 2. In order to study the effect of a slew on the oscillator performance, we have added the possibility of including a slew in the beam centroid at the initialization of MINERVA. Since this is a first attempt to simulate a slew in the simulation of multi-pass oscillators and since these slew measurements also represent the first such studies, we consider a linear slew in the simulations. It is not difficult to simulate more complex slew patterns, and we reserve this for future simulations for simplicity. In order to engage this option, we specified (1) the slew direction, (2) the number of bunches (one bunch for each pass) over which the slew will be applied, (3) the maximum slew displacement over the number of bunches, and (4) the displacement (from the axis of symmetry) at the start of the simulation. Note that the default setting is that the initial displacement is zero. Therefore, if we specify that there will be 100 bunches and that the maximum displacement will be 100 µm, then the displacement will increase by 1 µm on each pass. If it is required that the simulation goes beyond 100 passes, then this increase in the slew will continue at this rate for each additional pass.

Figure 14 shows the out-coupled power vs pass assuming (1) that there is no initial slew on the electron beam (blue), and (2) that the total slew is 100 µm (red) over 100 bunches. A steady state is reached after about 80 passes as in Fig. 4 at an output power level of about 8 MW in the absence of slew, while the power at the undulator exit is about 161 MW for an out-coupling of about 5%. The single-pass gain in the steady state is about 65%. The output power sags to about 5.5 MW with the slew at the end of 100 passes under these conditions with initial displacement of zero in the *x*-plane.

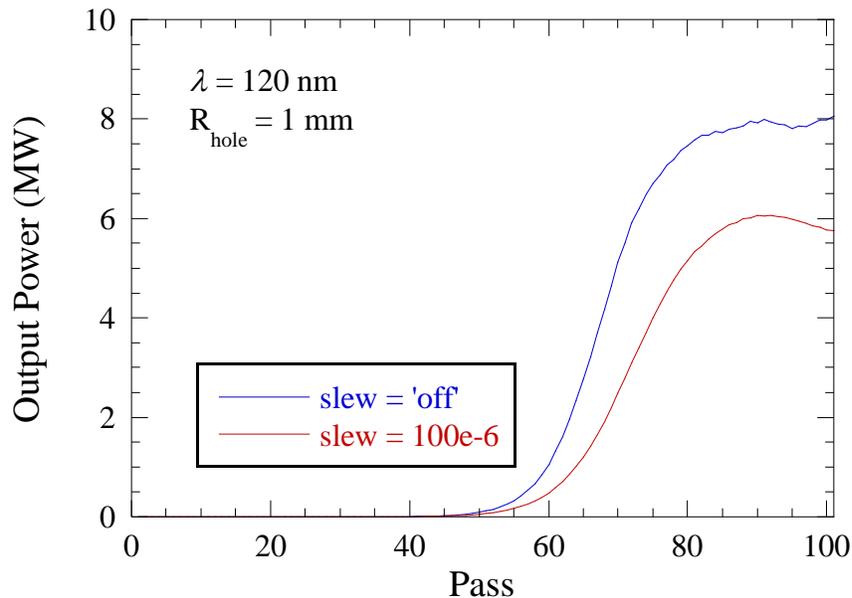

Fig. 14: Plot of output power vs pass number at 120 nm with no slew (blue curve) and with the 1-µm slew per bunch rate (red curve).

Figure 15 shows the effect of displacements (after 100 passes) in the *x*-direction which corresponds to the wiggler plane, and in the *y*-direction for slews which are comparable in magnitude to the beam dimensions. It is clear from the figure that the effect of the slew is greater if it is in the wiggle plane for this configuration. This is probably because there is no focusing in that direction in a planar undulator so the beam will remain off axis in that direction. While it might still excite radiation, the coupling to the optical modes will be reduced and this may lead to reduced out-coupling through the hole. It is also important to note that the degradation is only about 30% as the slew increases to 100 µm at the end of the simulation for a beam whose extent is about 120 µm. A mitigation effect occurs if the beam can be steered so it starts off axis at -1/2 the total slew in the *x*-plane.

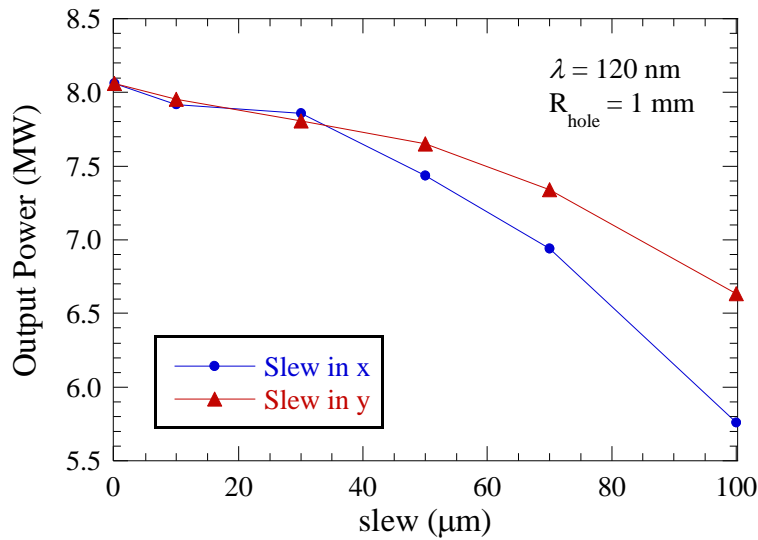

Fig. 15: A comparison of the effects of slew in the *x*-(or wiggle) plane (blue) and the *y*-plane (red) at 120 nm using MINERVA/OPC.

The effect of the slew on the performance of the aforementioned 13.4-nm oscillator case is shown in Fig. 16. Since about 250 passes were required to reach the steady state, we considered the maximum slew to be applied over that number of passes (bunches); hence, if the maximum slew is 10 μm over 250 bunches (passes) then a cumulative slew of 0.04 μm will be applied for each pass. As shown in the figure, the output power degrades by almost 80% as the maximum slew increases to 20 μm whether the slew is imposed in the *x*- or *y*-direction. This is much more sensitive than was found for the 120-nm oscillator, and we attribute this to the increased sensitivity to the relatively lower gain/pass possible at this shorter wavelength.

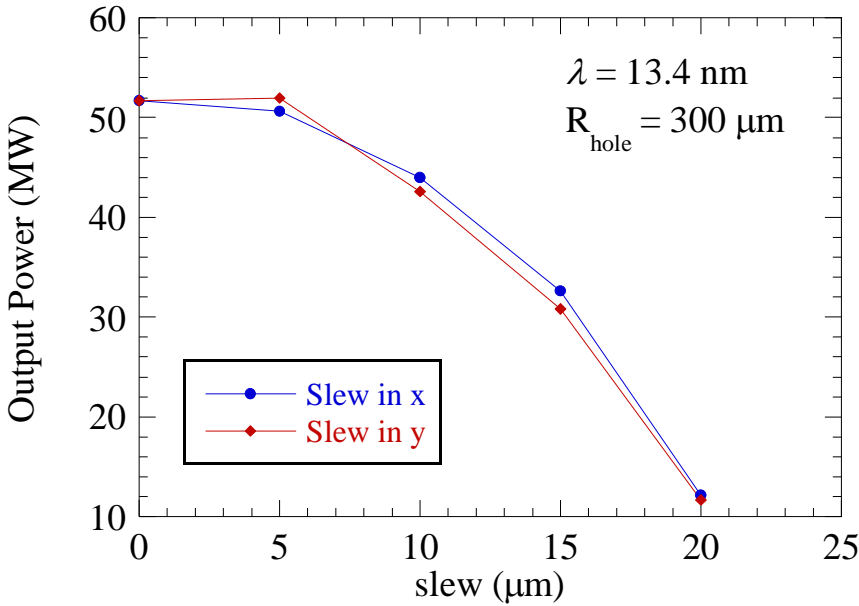

Fig. 16: A comparison of the effects of slew in the *x*-(or wiggle) plane (blue) and the *y*-plane (red) at 13.4 nm using MINERVA/OPC.

## VII. SUMMARY

In summary, we have described the proposed application at FAST of a 5-cm period undulator for the basis of unique tests of VUV-EUV FEL oscillator configurations. Simulations of the performance using GINGER and MINERVA/OPC at 120, 100, and 13.4 nm are very encouraging indicating saturation of the FELO with reasonable projected electron beam parameters and undulator length. We added consideration of the resonator mirror damage thresholds and refer to previous empirical results with SASE FEL pulses that show FELO energy densities are below the damage thresholds of the mirrors tested, particularly for the 120-nm case. In addition, we have considered for the first time the effects on the FELO performance at 120 nm and 13.4 nm should submacropulse centroid slews occur as may be driven by the HOMs in the TESLA-type cavities. Proper steering into the cavities to minimize the HOMs and at the undulator to make the slew symmetric in x should preserve FELO performance. These wakefield considerations would also apply to the recently proposed Tapering Enhanced Super-Radiant Stimulated Amplification (TESSA) based oscillator experiment at FAST at 515 nm [61] as well as the x-ray FELOs proposed to be driven with such TESLA cavities at LCLS-II [25] and the European XFEL [62].

Superconducting rf linac-based FELOs could provide a coherent photon source with wavelength extension far beyond the storage-ring based sources. There would be a concomitant potential for a myriad of science applications including use of the shortest wavelengths for tests of semiconductor industry interest in lithography. This potentially transform-limited source would have spectral purity without the spikiness of the SASE FELs and about 150 times more micropulses per second than those driven by present normal conducting rf linacs. Development of such unprecedented FELO sources is encouraged.

## VIII. ACKNOWLEDGMENTS

The FNAL author acknowledges earlier discussions with M. Church, N. Eddy, V. Shiltsev, S. Nagaitsev of Fermilab on ASTA/FAST, discussions with W. Colson of NPS on FELOs, the support of K. Robinson (LBNL), and the efforts of J. Santucci (FNAL) and A. Salehi (LBNL) in the shipping of the U5.0. This manuscript has been authored by Fermi Research Alliance, LLC under Contract No. DE-AC02-07CH11359 with the U.S. Department of Energy, Office of Science, Office of High Energy Physics.